\documentclass[prb,preprint]{revtex4-1}

\usepackage[x11names]{xcolor}
\usepackage{amsmath} 
\usepackage{amsfonts} 
\usepackage{graphicx} 
\usepackage{tikz}
\usepackage{tikzsymbols}
\usetikzlibrary{snakes}
\usetikzlibrary{decorations.pathmorphing,math,calc,3d,shapes}
\usetikzlibrary{arrows.meta,arrows,angles,quotes}
\usepackage{tikz-3dplot}
\usepackage{pgfplots}
\usepackage{mathtools}
\def\centerarc[#1](#2)(#3:#4:#5){
    \draw[#1]([shift=(#3:#5)]#2) arc (#3:#4:#5);
    }

\newcommand\be{\begin{equation}}
\newcommand\ee{\end{equation}}
\newcommand\nono{\nonumber}
\newcommand\bse{\begin{subequations}}
\newcommand\ese{\end{subequations}}

\begin{document}
\title{Singular Lagrangians and the Dirac--Bergmann Algorithm in Classical Mechanics}
\author{J. David Brown}
\email{david\_brown@ncsu.edu}
\affiliation{Department of Physics, North Carolina State University, Raleigh, NC 27695}
\date{\today}

\begin{abstract}  
Textbook treatments of  classical mechanics typically assume that the Lagrangian is nonsingular; that is, the matrix of second derivatives of the Lagrangian with respect to the velocities 
is invertible. This assumption insures that (i) Lagrange's equations can be solved for the accelerations as functions of coordinates and velocities, and (ii) 
the definitions of the conjugate momenta can be inverted to solve for the velocities as functions of coordinates and momenta. This assumption, however,  is unnecessarily restrictive---there are 
interesting classical dynamical systems with singular Lagrangians. 
The  algorithm for analyzing such systems was developed by Dirac and Bergmann in the 1950's. After a brief review of the Dirac--Bergmann algorithm, several 
 examples are presented using familiar components: point masses connected by massless springs, rods, cords and pulleys. 
\end{abstract}
%%%%%%%%%%%%%%%%%%%%%%%%%%%%%%%%%%%%%%%%%%%%%%%%%%%%%%%%%%%%%%%%%%%%%%%%%%%%%
\maketitle
%%\tableofcontents

%%%%%%%%%%%%%%%%%%%%%%%%%%
\section{Introduction}\label{sec:introduction}
The central focus of any advanced book on classical mechanics is the Lagrangian formulation of dynamics. With few exceptions these 
books assume that the Lagrangian $L(q,\dot q)$ is nonsingular. That is, the matrix 
\be
	L_{ij} \equiv \frac{\partial^2 L}{\partial \dot q_i \partial \dot q_j}
\ee
of second derivatives of $L$ with respect to the velocities $\dot q_i \equiv dq_i/dt$  is invertible. 
If $L_{ij}$ is not invertible, we cannot solve Lagrange's equations for the accelerations as functions of the coordinates and velocities. 

The starting point for the Hamiltonian formulation of mechanics is the Lagrangian.  If $L_{ij}$  is not invertible, 
the definitions of momenta in terms of coordinates and velocities cannot be inverted for the velocities as functions of coordinates and momenta.
The Hamiltonian theory cannot be constructed in the usual way. 

The purpose of this article is to point out that there are, in fact, physically interesting classical dynamical systems with singular Lagrangians. The formalism for treating such systems was developed in the 1950's by Dirac \cite{Dirac1950,Dirac1951,Dirac1958a,Dirac1958b,Dirac1959} and by Bergmann and collaborators,\cite{Bergmann1949,BergmannBrunings1949,BergmannEtAl1950,AndersonBergmann1951,Penfeld1951,BergmannGoldberg1955} following earlier 
work by Rosenfeld.\cite{RosenfeldGerman}  This formalism is referred to as the Dirac--Bergmann algorithm. To be more precise, Dirac and Bergmann, and also 
Rosenfeld,\cite{Salisbury:2016xdf}  showed that a singular Lagrangian system can be placed in the form of a ``constrained Hamiltonian system" in which the 
evolution is constrained to a subspace 
of phase space.  The singular nature of the system is most clearly exhibited in Hamiltonian form. 

The primary motivation for Dirac and Bergmann was to understand the structure of field theories such as electromagnetism and general relativity.  These theories, as well as 
Yang--Mills theory and string theory, are gauge theories---they contain degrees of freedom that do not alter the physical state of the system. Gauge theories are described by singular Lagrangians, but (as will be seen in the examples) 
not all singular Lagrangian systems are gauge theories. 

In Section \ref{sec:recipe}  we outline the Dirac--Bergmann algorithm, the steps for converting a singular Lagrangian system into a constrained Hamiltonian system. The detailed reasoning is spelled out in numerous 
books \cite{DiracLectures, SudarshanMukunda, HansonReggeTeitelboim, Sundermeyer, HenneauxTeitelboim, RotheRothe, Deriglazov, Lusanna:2019cmq} and review articles.\cite{Date:2010xr, Salisbury:2016xdf, Brown:2022gha} 
Section \ref{sec:examples}  contains  a number of physical examples of singular systems constructed from familiar elements found in textbook classical mechanics problems: point masses connected by massless 
springs, rods, cords and pulleys.  

In Sec.~\ref{sec:compoundspring} we consider a ``compound spring" obtained by welding a spring with stiffness $k_1$ and relaxed length $\ell_1$ to a second spring with stiffness $k_2$ and relaxed length $\ell_2$.   The system of Sec.~\ref{sec:pendulumtwosprings}  consists of a pendulum attached to two spring.,\cite{CisnerosParra}
In Sec.~\ref{sec:massesspringsrings} we analyze three masses moving on a circular ring and  connected by springs.   None of these first  three examples contain any gauge freedom. 

In Sec.~\ref{sec:massesrodssprings}  we consider four masses fixed at the centers of freely extensible rods. The ends of the rods are connected, and the connection points (the ``corners")  slide freely on the vertical posts. We consider two versions of this system: one with  springs extending from  the corners to the ceiling, the other with  springs extending from the masses to the ceiling.  In both cases the system is described by a singular Lagrangian, but the placement of the springs plays an important role which is clearly revealed in the Hamiltonian formulation.  When the springs are attached to the masses, the system contains gauge freedom. When the springs are attached to the corners, there is no gauge freedom. 

The system discussed in Sec.~\ref{sec:pairsofpulleys} consists of a single loop of cord weaving between three pairs of pulleys. 
The lower pulley of each pair is fixed, while the upper pulley is attached to a mass and a spring. 
This system is a gauge theory, and can be generalized to any number of pairs of pulleys. 

Section \ref{section:more} lists two more problems described by singular Lagrangians, without solutions.  These problems are left as exercises for the reader. 

The common element in each of these singular systems is the presence of degrees of freedom with no inertial response. Consider, for example, a two--particle 
system with coordinates $x_1$, $x_2$, and Lagrangian $L = m_1 \dot x_1^2/2 + m_2 \dot x_2^2/2 - V(x_1,x_2)$. The matrix of second derivatives of $L$ is  
nonsingular since $L_{ij}$ is diagonal with entries $m_1$ and $m_2$. Now 
set the mass $m_1$ to zero so that $L$ becomes singular.  We can now vary the coordinate $x_1$ without any inertial response---changing $x_1$ does not cause any mass in the system  to move. 
This observation provides an intuitive test for singular systems in classical mechanics. Imagine fixing the location of each mass, and ask: Is the system rigid? 
If not, then there are degrees of freedom with no inertial response.  The Lagrangian for such a system is singular. 

Section  \ref{sec:conclude} contains concluding remarks. 

%%%%%%%%%%%%%%%%%%%%%%%%%%%%%%%%%%%%%%
\section{Dirac--Bergmann Algorithm}\label{sec:recipe}
Consider a system with $\bar N$ generalized coordinates $q_i$, where $i=1,\ldots,\bar N$. The velocities are denoted by $\dot q_i$. The Lagrangian
$L(q,\dot q)$ is 
assumed to be singular, so the rank of the matrix $L_{ij}$ (the number of linearly independent rows or columns) 
is less than $\bar N$, say, $\bar M$. We assume that the rank $\bar M$ is constant throughout phase space. The following is a short summary of the Dirac--Bergmann algorithm for converting this system into constrained Hamiltonian form. This summary is not intended as a substitute for the more thorough treatments given elsewhere.\cite{DiracLectures,SudarshanMukunda, HansonReggeTeitelboim, Sundermeyer,HenneauxTeitelboim, RotheRothe,Deriglazov, Lusanna:2019cmq, Date:2010xr, Salisbury:2016xdf, Brown:2022gha} 

\begin{itemize}
\item Compute the conjugate momenta $p_i = \partial L/\partial \dot q_i$. Since the Lagrangian is singular, these relations cannot be inverted 
for the velocities as functions of coordinates and momenta. This implies the existence of  $\bar N-\bar M$ relations among the 
coordinates and momenta.\cite{Lquadratic}  These relations are the {\em primary constraints}, denoted $\phi_a(q,p) = 0$, with the index $a$ 
ranging from $1$ to $\bar N-\bar M$.
\item Define the {\em canonical Hamiltonian} $H_C$ by writing  $p_i\dot q_i - L(q,\dot q)$  in terms of the $q$'s and $p$'s. It can be shown that 
this is always possible. Note that $H_C(q,p)$ is not unique, because one can always use the constraints $\phi_a(q,p) =0$ to write some of the canonical variables in terms of  others. 
\item Define the {\em primary Hamiltonian} $H_P$ by adding the primary constraints with Lagrange multipliers to the canonical Hamiltonian. That is, 
$H_P  = H_C + \lambda^a \phi_a$, where $\lambda_a$ are the Lagrange multipliers. 
\item Impose the conditions $[\phi_a,H_P] = 0$, referred to as ``consistency conditions,"  where $[\, , \, ]$ is the Poisson bracket.  These conditions insure that the primary constraints 
are preserved under time evolution. 
The consistency  conditions (one for each value of the index $a$) will reduce to a combination of (i) identities when the primary constraints $\phi_a(q,p) = 0$ hold; (ii) restrictions on the Lagrange multipliers; and/or (iii) restrictions on the $q$'s and $p$'s. The restrictions on the Lagrange multipliers express some $\lambda$'s in terms of 
$q$'s, $p$'s, and the remaining $\lambda$'s. 
Restrictions on the $q$'s and $p$'s are  {\em secondary constraints}, which we write as $\psi_m(q,p) = 0$. 
\item The consistency conditions must be  applied to the secondary constraints  to insure their preservation in time: $[\psi_m,H_P] = 0$.  This can yield identities, 
further restrictions on the Lagrange multipliers, and/or  {\em tertiary constraints}, which are further restrictions on the $q$'s and 
$p$'s. We continue to apply the consistency conditions to identify higher--order constraints and restrictions on the Lagrange multipliers. The process naturally stops when the consistency conditions have been applied to all constraints.  We extend the range of the index $m$ and 
let $\psi_m(q,p)$ denote all of the secondary, tertiary, and higher--order constraints.  
\item The {\em total Hamiltonian} $H_T$  is obtained from the primary Hamiltonian $H_P$ by incorporating the restrictions 
on Lagrange multipliers. In the most general case, a subset of the Lagrange multipliers will remain free. 
\item The primary, secondary, tertiary, {\em etc.} constraints 
$\phi_a$ and $\psi_m$ are separated into first and second class. First class constraints have the 
property that their Poisson bracket with {\em all} constraints vanish when the constraints hold. Second class constraints have nonvanishing 
Poisson bracket with at least one other constraint. Let ${\cal C}^{(fc)}_\alpha$ denote the set of first class constraints, and ${\cal C}^{(sc)}_\mu$ denote the set of second class constraints. 
\item  A subset of first class constraints can be constructed from the primary constraints $\phi_a(q,p)$. These are the 
{\em primary first class constraints} which we denote ${\cal C}^{(pfc)}_A$. It can be shown that the total Hamiltonian can be written as 
$H_T = H_{fc} + \Lambda^A {\cal C}_A^{(pfc)}$, where  the Lagrange multipliers $\Lambda^A$ are free and the 
{\em first class Hamiltonian} $H_{fc}$ has vanishing Poisson bracket with all of the constraints (when the constraints hold). 
\end{itemize}

Before continuing, a few comments are in order.  The equations of motion generated by the total Hamiltonian $H_T$ through the Poisson bracket are equivalent to  
Lagrange's equations for the original Lagrangian system. 
Since the Lagrange multipliers $\Lambda^A$ are completely arbitrary, the phase space transformations generated by   
the primary first class constraints ${\cal C}^{(pfc)}_A$  do not change the physical state of the system.  We refer to such  transformations as 
{\em gauge transformations}.\cite{gaugecomment} 
Therefore, primary first  class constraints generate gauge transformations.  

The Dirac conjecture\cite{DiracLectures} says that {\em all} first class constraints ${\cal C}^{(fc)}_\alpha$ 
generate gauge transformations.  Counterexamples to this conjecture have been described in the literature by 
a number of researchers.\cite{HenneauxTeitelboim, Cawley, Frenkel, SuganoKimura, Li, Wu, MiskovicZanelli, WangLiWang} Other researchers  
have argued against these counterexamples, citing subtleties in the way that the constraints are written.\cite{RotheRothe, RotheRothe2} 
The Dirac conjecture is often taken as an assumption.\cite{Diracconjecturecomment}

Here is the next step in the algorithm: 
\begin{itemize} 
\item Assuming the Dirac conjecture holds, each of the first class constraints has the status of a gauge generator. These 
constraints can be treated on an equal footing by constructing 
the {\em extended Hamiltonian}  $H_E = H_{fc} + \Lambda^\alpha {\cal C}^{(fc)}_\alpha$. This is the sum of the first class Hamiltonian $H_{fc}$ 
and a linear combination of all first class constraints with unrestricted Lagrange multipliers $\Lambda^\alpha$. The equations of motion defined by the extended 
Hamiltonian are not strictly equivalent to the original Lagrangian equations of motion. Nevertheless, the theories agree for the evolution of physical variables 
(variables that are invariant under gauge transformations.)  
\end{itemize}

Phase space functions $F$ are evolved in time with either the extended Hamiltonian, $\dot F = [F,H_E]$,  or the total Hamiltonian, $\dot F = [F,H_T]$.  
Physical trajectories are those that lie in the 
subspace of phase space where the constraints hold. 

The constraint relations can be used 
freely after computing Poisson brackets, but not before.  
For example, the constraints can be used to alter the equations of motion $\dot F = [F,H_E]$  (or $\dot F = [F,H_T]$) but not the functions that appears in the Poisson bracket. 

We now have  options. One option:
\begin{itemize}
\item Eliminate the second class constraints leaving the gauge freedom generated by the first class constraints intact. 
We do this by replacing the Poisson bracket with the Dirac bracket, defined as follows. Let 
${\cal M}_{\mu\nu} = [{\cal C}_\mu^{(sc)}, {\cal C}_\nu^{(sc)} ]$ 
denote the matrix of Poisson brackets among the second class constraints, and let ${\cal M}^{\mu\nu}$ denote its inverse. The Dirac bracket 
is  
\be\label{DiracBracketsSC}
	[F,G]^* = [F,G] - [F,{\cal C}^{(sc)}_\mu] {\cal M}^{\mu\nu} [{\cal C}^{(sc)}_\nu,G]
\ee
where $F$ and $G$ are phase space functions. (Summation over repeated indices is implied.) 
\end{itemize}

Like the Poisson bracket, the Dirac bracket is antisymmetric and obeys the Jacobi identity.  It also satisfies  $[F,{\cal C}^{(sc)}_\mu]^* = 0$ for any phase space function $F$. 
This allows us to use the second class constraints to simplify  $F$ and $G$  {\em before} computing the bracket  $[F,G]^*$. 

Because the Poisson bracket of the extended Hamiltonian with a second class constraint will vanish when the constraints hold, it follows that 
$[F,H_E]^*$ equals $[F,H_E]$   when the constraints hold. (Likewise for the total Hamiltonian $H_T$.) Thus, the equations of motion  can be 
defined using either the Dirac bracket or the Poisson bracket. 
 
\begin{itemize}
\item We can now eliminate a subset of phase space variables by imposing the second class constraints ${\cal C}^{(sc)}_\mu = 0$ and using the Dirac bracket. 
In particular, we can use  ${\cal C}^{(sc)}_\mu = 0$ to eliminate variables from the extended Hamiltonian (or total Hamiltonian), resulting in a {\em partially 
reduced Hamiltonian} $H_{PR}$.   Time evolution becomes  $\dot F = [F,H_{PR}]^*$. 
\end{itemize}

A second option: 
\begin{itemize}
\item Eliminate both first and second class constraints by imposing gauge conditions. Canonical gauge conditions, like constraints, are 
restrictions on  the phase space variables. Let us denote the constraints and gauge conditions, combined, by ${\cal C}^{(all)}_M$. A good set of 
gauge conditions will have the property that ${\cal C}^{(all)}_M$ are second class. That is, the matrix of Poisson brackets 
${\cal M}_{MN} = [{\cal C}_M^{(all)}, {\cal C}_N^{(all)} ]$
is invertible. Let ${\cal M}^{MN}$ denote the inverse and define the Dirac bracket by
\be\label{DiracBracketsALL}
	[F,G]^* = [F,G] - [F,{\cal C}^{(all)}_M] {\cal M}^{MN} [{\cal C}^{(all)}_N,G] \ ,
\ee
where summations over $M$ and $N$ are implied. 
\item Now use the constraints and gauge conditions ${\cal C}^{(all)}_M = 0$  
 to eliminate a subset of phase space variables. We can eliminate variables 
 from the extended Hamiltonian (or total Hamiltonian), resulting in a {\em fully reduced Hamiltonian} $H_{FR}$. Time evolution becomes $\dot F = [F,H_{FR}]^*$. 
\end{itemize}

In the next section we apply the Dirac--Bergmann algorithm to analyze problems in classical mechanics that are described by singular Lagrangians. 

%%%%%%%%%%%%%%%%%%%%%%%%%%%%%%%%%%%%%%%%%
\section{Examples}\label{sec:examples}
\subsection{Compound spring}\label{sec:compoundspring}
\begin{figure}[htb] 
\centering
\begin{tikzpicture}[]
    \draw [draw=black, thick, fill=Bisque3] (-2,0)--(2,0)--(2,0.4)--(-2,0.4)--cycle;
    % springs
    \draw[Blue3, snake=coil,segment amplitude=5pt, segment aspect=1, segment length=9, thick] (0,0) -- (0,-2);
    \node at (0.8,-1) {$k_1$, $\ell_1$};
    \draw[Blue3, snake=coil,segment amplitude=4pt, segment aspect=0.9, segment length=8, very thick] (0,-2) -- (0,-4);
     \node at (0.8,-3) {$k_2$, $\ell_2$};
     \draw [very thick] (0,-1.78)--(0,-2.01);
    % mass
    \draw [draw=black, thick,  fill=Firebrick3] (0,-4) circle (0.25);
    \node at (0.6,-4) {$m$};
    \draw [->] (-0.6,0)--(-0.6,-1.9) node[midway, left] {$x_1$};
    \draw [->] (-0.6,-2.0)--(-0.6,-3.95) node[midway, left] {$x_2$};
\end{tikzpicture}
\caption{The compound spring. The mass moves vertically,  with gravity acting in the downward direction. The generalized coordinates are 
the lengths of the two springs. }
\label{fig:compoundspring}
\end{figure}
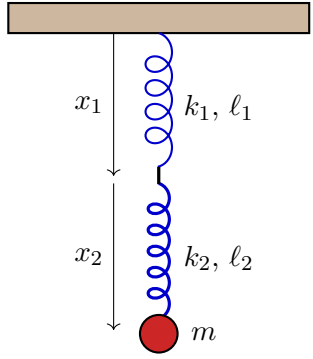

Form a ``compound spring" by welding two springs together, as shown in Fig.~\ref{fig:compoundspring}.  
One end of the compound spring is attached to the ceiling, and a mass $m$ hangs from the other end. Let $x_1$ and $x_2$ denote the lengths of the two springs,  so the distance between the ceiling and the mass is $x_1 + x_2$.  The  Lagrangian for this system is
\be
	L  = \frac{m}{2} (\dot x_1 + \dot x_2)^2  + m g (x_1 + x_2) - \frac{k_1}{2} (x_1 - \ell_1)^2 - \frac{k_2}{2} (x_2 - \ell_2)^2 \ .
\ee
The matrix of second derivatives of $L$ with respect to the velocities $\dot x_i$, 
\be\label{compoundspringhessian}
	L_{ij} = \begin{pmatrix} m & m \\ m & m \end{pmatrix} \ ,
\ee
is singular with rank 1. The momenta are 
\bse
\begin{align}
	p_1 & = \frac{\partial L}{\partial \dot x_1} = m(\dot x_1 + \dot x_2)  \ , \\
	p_2 & = \frac{\partial L}{\partial \dot x_2} = m(\dot x_1 + \dot x_2) \ ,
\end{align}
\ese
and we can identify the primary constraint 
\be
	\phi \equiv p_2 - p_1 
\ee
by inspection. 

Next, construct the canonical Hamiltonian by writing $p_i \dot x_i - L$ (a sum over the repeated index $i$ is implied) in terms of $p$'s and $x$'s: 
\be
	H_C(x,p) = \frac{1}{2m}  p_1 p_2  - m g (x_1 + x_2) + \frac{k_1}{2} (x_1 - \ell_1)^2 + \frac{k_2}{2} (x_2 - \ell_2)^2 \ .
\ee
The leading term $p_1 p_2/(2m)$ can be written in other ways, such as $p_1^2/(2m)$ or $(p_1^2 + p_2^2)/(4m)$, by invoking the 
constraint $\phi = 0$. 

The primary Hamiltonian is $H_P = H_C + \lambda \phi$. The consistency condition $[\phi,H_P] = 0$ yields the secondary constraint
\be
	\psi = k_1(x_1 - \ell_1) - k_2(x_2 - \ell_2) \ ,
\ee
and the  condition $[\psi,H_P] = 0$ restricts the Lagrange multiplier to
\be\label{compoundspringlambda}
	\lambda = \frac{k_1 p_2 - k_2 p_1}{2m(k_1 + k_2)}  \ .
\ee
The application of consistency conditions is now complete. 

The secondary constraint $\psi=0$ has a direct physical interpretation via Newton's third law. It tells us that the force  $k_1(x_1 - \ell_1)$ that spring 1 exerts on spring 2 is equal but opposite to the 
force $k_2(\ell_2 - x_2)$ that spring 2 exerts on spring 1.

The total Hamiltonian is obtained by using the result (\ref{compoundspringlambda}) for $\lambda$ in the primary Hamiltonian: 
\be
	H_T =  \frac{1}{2m}  p_1 p_2  +  \frac{k_1 p_2 - k_2 p_1}{2m(k_1 + k_2)} (p_2 - p_1) 
	- m g (x_1 + x_2) + \frac{k_1}{2} (x_1 - \ell_1)^2 + \frac{k_2}{2} (x_2 - \ell_2)^2 \ .
\ee
The two constraints are second class, since $[\phi,\psi] = k_1 + k_2$ is nonzero. There are no first class constraints, so the system has no gauge freedom and the  
total Hamiltonian, first class Hamiltonian, and extended Hamiltonian coincide: $H_T = H_{fc} = H_E$. 

Let  ${\cal C}^{(sc)}_\mu = \{ \phi, \psi \}$ denote the set of second class constraints.  The matrix ${\cal M}_{\mu\nu}  = [{\cal C}^{(sc)}_\mu, {\cal C}^{(sc)}_\nu] $ 
is invertible with inverse
\be
	{\cal M}^{\mu\nu}  = \frac{1}{k_1 + k_2}  \begin{pmatrix}  0 & -1\\ 1 & 0   \end{pmatrix} \ .
\ee
We now construct the Dirac bracket as in Eq.~(\ref{DiracBracketsSC}). The nonzero brackets among the phase space variables are 
\bse
\begin{align}
	[x_1,p_1]^* & = [x_1, p_2]^*  = k_2/(k_1 + k_2)  \ , \\
	 [x_2,p_1]^* & = [x_2,p_2]^*  = k_1/(k_1 + k_2)  \ .
\end{align}
\ese
We can use the constraints to eliminate two of the phase space variables. For example,  solving $\phi = \psi = 0$ for $x_2$ and $p_2$, we find 
\bse\label{compoundspringx2p2soln}
\begin{align}
	x_2 & =  \ell_2 +  \frac{k_1}{k_2} (x_1 - \ell_1)   \ , \\
	p_2 & = p_1  \ ,
\end{align}
\ese
and  the Hamiltonian reduces to 
\be
	H_R = \frac{p_1^2}{2m}  - \frac{mg}{k_2} \bigl[ (k_1 + k_2)x_1 + k_2 \ell_2 - k_1 \ell_1 \bigr]  + \frac{1}{2} ( k_1 + k_1^2/k_2) (x_1 - \ell_1)^2 \ .
\ee
Note that in the absence of first class constraints, the partially and fully reduced Hamiltonians coincide. Here we use $H_R$ 
to denote this reduced Hamiltonian. 

The time evolution of any function of the phase space variables $x_1$, $p_1$, $x_2$, $p_2$ can be obtained from $H_R$ and the Dirac bracket. In particular we have
\bse
\begin{align}
	\dot x_1 & = [x_1,H_R]^* =  \frac{k_2}{m(k_1 + k_2)} p_1 \ , \\
	\dot p_1 & = [p_1,H_R]^* = -k_1 (x_1 - \ell_1) + mg \ ,
\end{align}
\ese
which form a closed set of differential equations for $x_1$ and $p_1$ with general solution 
\bse
\begin{align}
	x_1(t)  & = A\cos(\omega t) + B\sin(\omega t) + \ell_1 + mg/k_1 \ , \\
	p_1(t) & = \frac{k_1}{\omega} \left( B\cos(\omega t) - A\sin(\omega t) \right) \ .
\end{align}
\ese
Here,  $A$ and $B$ are constants and the angular frequency is defined by 
\be
	\omega = \sqrt{ k_1 k_2/(m(k_1 + k_2)) }  \ .
\ee
Given $x_1(t)$, we can determine $x_2$ 
as a function of time from the result (\ref{compoundspringx2p2soln}a). The position of the mass below the ceiling, $x_1(t) + x_2(t)$, then follows. We find that the  mass
executes simple harmonic motion about its equilibrium position $\ell_1 + \ell_2 + mg(k_1 + k_2)/(k_1 k_2)$ with angular frequency $\omega$.

%%%%%%%%%%%%%%%%%%%%%%%%%%%%%%%%%%%%%%%%%%
\subsection{Pendulum and two springs}\label{sec:pendulumtwosprings}

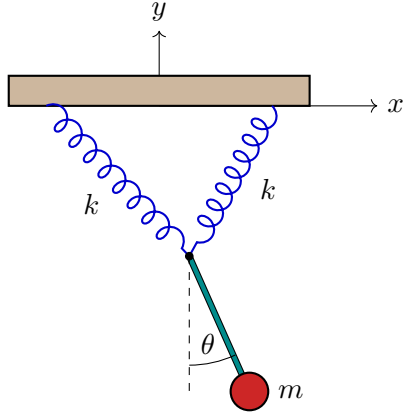
\begin{figure}[htb] 
\centering
\begin{tikzpicture}[]
      \draw [draw=black, thick,  fill=Bisque3] (-2,0)--(2,0)--(2,0.4)--(-2,0.4)--cycle;
      \draw [->] (0,0)--(2.9,0) node[right] {$x$}; 
      \draw [->] (0,0.4)--(0,1) node[above] {$y$}; 
    % springs
    \draw[Blue3, snake=coil,segment amplitude=4pt, segment aspect=0.8, segment length=8, thick] (-1.5,0) -- (0.4,-2);
    \draw[Blue3, snake=coil,segment amplitude=4pt, segment aspect=0.8, segment length=7, thick] (1.5,0) -- (0.4,-2);
    \node at (-0.9,-1.3) {$k$};
    \node at (1.45,-1.1) {$k$}; 
    % pendulum
    \draw [double=Cyan4, double distance=1.7pt] (0.4,-2) -- (1.2,-3.8);
    \draw [draw=black, thick,  fill=Firebrick3] (1.2,-3.8) circle (0.25);
    \node at (1.75, -3.8) {$m$};
    \draw [fill] (0.4,-2) circle (0.05);
    \draw [dashed] (0.4,-2)--(0.4,-3.8); 
    \node at (0.65,-3.17) {$\theta$};
    \centerarc[](0.4,-2)(-90:-65:1.45);
\end{tikzpicture}
\caption{A pendulum hanging from two springs. The springs are attached to the ceiling at the points $x=\pm d$, $y=0$. }
\label{fig:pendulumtwosprings}
\end{figure}

Figure \ref{fig:pendulumtwosprings} shows a pendulum of length $\ell$ hanging from two springs. Each spring has stiffness $k$, and for simplicity we take the 
relaxed length of each spring to be zero. The generalized coordinates 
are the Cartesian coordinates $x$ and $y$ of the point where the springs attach to the pendulum, and the angle 
$\theta$ of the pendulum rod. (The Cartesian coordinate origin is midway between the points where the springs attach to the ceiling. 
The angle $\theta$ is measured from the negative $y$--axis.) 

The kinetic energy for this system is $T = (m/2)(\dot X^2 + \dot Y^2)$, where $X = x + \ell\sin\theta$ and 
$Y = y - \ell\cos\theta$ are the Cartesian coordinates of the mass $m$.  The Lagrangian is
\be\label{pendulumtwospringsLag}
	L = \frac{m}{2} (\dot x^2 + \dot y^2  + \ell^2 \dot\theta^2) + m\ell(\dot x \cos\theta +\dot y  \sin\theta)\dot\theta - mg( y - \ell\cos\theta ) 
	- k(x^2 + y^2 + d^2) \ ,
\ee
and the matrix of second derivatives  of $L$ with respect to the velocities $\dot x$, $\dot y$, $\dot\theta$ is
\be
	L_{ij} = \begin{pmatrix}  m & 0 & m\ell\cos\theta \\ 0 & m & m\ell\sin\theta \\ m\ell\cos\theta & m\ell\sin\theta & m\ell^2 \end{pmatrix} \ .
\ee
This matrix is singular with rank $2$. 

The momenta for this system are 
\bse
\begin{align}
	p_x & =  m\dot x + m\ell \dot\theta  \cos\theta  \ ,\\
	p_y & = m\dot y + m\ell \dot\theta \sin\theta  \ ,\\
	p_\theta & = m\ell (\dot x \cos\theta + \dot y\sin\theta ) + m\ell^2\dot \theta \ .
\end{align}
\ese
Since the Lagrangian is quadratic in the velocities, the constraint can be constructed as $\phi = V^i (p_i - L_i)$,  where   the vector $V^i = (-\ell\cos\theta, -\ell\sin\theta, 1)$ spans 
the null space of  $L_{ij}$. (Details are given at the end of this paper.\cite{Lquadratic})  In this case $L_i = 0$ and the primary constraint is
\be
	\phi  = - \ell p_x\cos\theta - \ell p_y\sin\theta + p_\theta \ .
\ee
The canonical Hamiltonian can be written as 
\be
	H_C = \frac{1}{2m}(p_x^2 + p_y^2)  + k(x^2 + y^2 + d^2)  + mg(y- \ell\cos\theta) \ ,
\ee
and the primary Hamiltonian is $H_P = H_C + \lambda\phi$. 

The consistency condition $[\phi,H_P] = 0$ yields the secondary constraint 
\be
	\psi = 2k\ell (x\cos\theta + y\sin\theta) \ ,
\ee
which gives $\tan\theta = -x/y$. This tells us that the angle of the force exerted by the springs on the massless connection point must coincide with the angle of the pendulum rod. This is a 
consequence of Newton's third law---the forces exerted by the springs on the rod must be equal in magnitude but opposite in direction to the force that the rod exerts on the springs---and 
the fact that the rod can only exert a force along its own direction, at angle $\theta$. 

The condition $[\psi,H_P] = 0$ determines the Lagrange multiplier to be 
\be\label{pendulumtwospringsLM}
	\lambda = \frac{p_x\cos\theta + p_y\sin\theta}{m(\ell + x\sin\theta - y\cos\theta)} \ ,
\ee
and
\begin{align}
	H_T = & \frac{1}{2m}(p_x^2 + p_y^2)  + k(x^2 + y^2 + d^2)  + mg(y- \ell\cos\theta)   \nono\\
	& +  \frac{p_x\cos\theta  + p_y\sin\theta}{m(\ell + x\sin\theta - y\cos\theta)} 
	(  - \ell p_x\cos\theta - \ell p_y\sin\theta + p_\theta  ) 
\end{align}
is the  total Hamiltonian. 

Note that the denominator in Eq.~(\ref{pendulumtwospringsLM})  is the coefficient of $\lambda$ in 
$[\psi,H_P]$. This coefficient vanishes when $x= - \ell\sin\theta$, $y = \ell\cos\theta$. At these points in phase space the Lagrange multiplier is not determined. 
This is not a shortcoming of the Dirac--Bergmann formalism, rather, it is a property of the physical system defined by the Lagrangian (\ref{pendulumtwospringsLag}). When 
$x= - \ell\sin\theta$ and $y = \ell\cos\theta$, the mass $m$ is at the origin and the pendulum rod can rotate without any inertial resistance, and without any change in the 
potential energy. Thus, at these points in phase space, the system exhibits a gauge--like freedom in which multiple configurations are physically indistinguishable. We can avoid 
this complication by assuming the mass stays below the ceiling, so that $Y = y - \ell\cos\theta$ is always negative. 

With this assumption  the constraints are second class: $[\phi,\psi] = 2k\ell ( \ell + x\sin\theta - y\cos\theta) \ne 0$. There is no gauge freedom, so $H_E = H_{fc} = H_T$. 
We can construct the Dirac bracket from Eq.~(\ref{DiracBracketsSC}) with 
\be
	{\cal M}^{\mu\nu} = \frac{1}{2 k \ell(\ell + x\sin\theta - y\cos\theta)} \begin{pmatrix} 0 & -1 \\ 1 & 0 \end{pmatrix} \ .
\ee
The constraints $\phi = \psi = 0$ imply\cite{pendulumtwospringscomment}
\bse\label{pendulumtwospringsthpth}
\begin{align}
	\theta & = -\arctan (x/y) \ ,\\
	p_\theta & = \frac{\ell}{\sqrt{x^2 + y^2}} (x p_y - y p_x) \ ,
\end{align}
\ese
and we can use these results to reduce the Hamiltonian: 
\be
	H_R = \frac{1}{2m} (p_x^2 + p_y^2) + m g y ( 1 + \ell/\sqrt{x^2 + y^2})  + k(x^2 + y^2 + d^2) \ .
\ee
The nonzero Dirac brackets among the remaining variables are
\bse
\begin{align}
	[x,p_x]^* & = \frac{r + \ell x^2/r^2}{r+\ell} \ ,\\
	[x,p_y]^* = [y,p_x]^*&  = \frac{\ell x y /r^2}{r+\ell} \ ,\\
	[y,p_y]^* & = \frac{r + \ell y^2/r^2}{r+\ell}
\end{align}
\ese
where $r \equiv \sqrt{x^2 +  y^2}$. 
The equations of motion are 
\bse
\begin{align}
	\dot x & = [x,H_R]^* = \frac{\ell x (x p_x + y p_y)  + p_x r^3 }{m r^2 ( r + \ell )}  \ ,\\
	\dot y & =  [y,H_R]^* = \frac{\ell y (x p_x + y p_y)  + p_y r^3 } {m r^2  (r + \ell )}  \ ,\\
	\dot p_x & = [p_x,H_R]^* = -2kx \ ,\\
	\dot p_y & = [p_y,H_R]^* = -mg - 2ky \ .
\end{align}
\ese
These can be solved numerically in a straightforward fashion. The angle $\theta(t)$ and its conjugate $p_\theta(t)$  follow from Eqs.~(\ref{pendulumtwospringsthpth}). 

%%%%%%%%%%%%%%%%%%%%%%%%%%%%%%%%%%%%%%%%%
\subsection{Masses, springs and ring}\label{sec:massesspringsrings}
\begin{figure}[htb]
\begin{tikzpicture}[]
    % ring
    \draw [double=Cyan4, double distance=2.0pt] circle (2.00);
    % masses
    \draw [draw=black, thick,  fill=Firebrick3] ({2*sin(30)},{2*cos(30)}) circle (0.25);
    \draw [draw=black, thick,  fill=Firebrick3] ({2*sin(150)},{2*cos(150)}) circle (0.25);
    \draw [draw=black, thick,  fill=Firebrick3] ({2*sin(300)},{2*cos(300)}) circle (0.25);
    % springs
    \tikzmath{\a1=30; \a2=160; \a3=300;}
    \draw[Blue3, snake=coil, segment amplitude=4pt, segment aspect=0.8, thick, segment length=6] (0.2,0.3) -- ({2*sin(\a1)},{2*cos(\a1)});
    \draw[Blue3, snake=coil, segment amplitude=4pt, segment aspect=0.8, thick, segment length=7] (0.2,0.3) -- ({2*sin(150)},{2*cos(150)});
    \draw[Blue3, snake=coil, segment amplitude=4pt, segment aspect=0.8, thick, segment length=8] (0.2,0.3) -- ({2*sin(300)},{2*cos(300)});
    \centerarc[->](0,0)(0:40:2.6);
    \draw [->] (2.3,0)--(3.2,0) node[right] {$x$}; 
    \draw [->] (0,2.2)--(0,3.1) node[right] {$y$};
    \node at ({2.6*sin(30)},{2.6*cos(30)}) {$\theta_1$};
    \node at ({2.6*sin(150)},{2.6*cos(150)}) {$\theta_3$};'
    \node at ({2.6*sin(300)},{2.6*cos(300)}) {$\theta_2$};
\end{tikzpicture}
\caption{Three masses connected by springs  move without friction on a circular ring.}
\label{fig:massesspringsrings}
\end{figure}
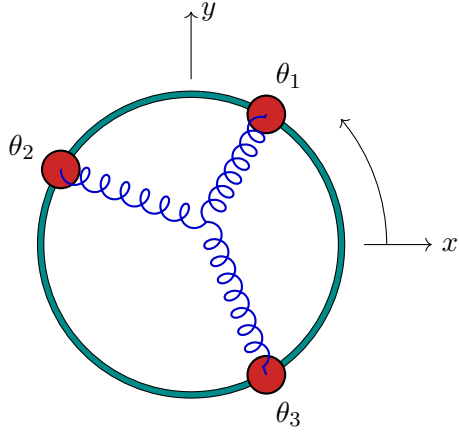

Three identical masses slide without friction on a ring of radius $R$. The masses are connected by springs, as shown in Fig.~\ref{fig:massesspringsrings}. Each spring 
has stiffness $k$ and for simplicity we set the relaxed lengths equal to zero. The generalized coordinates for this system are the angles 
$\theta_1$, $\theta_2$, $\theta_3$ of the three masses and the 
Cartesian coordinates $x$ and $y$ of the point where the springs connect. (The origin is at the center of the ring. All angles are measured with respect to the $x$--axis.) 

The Lagrangian for this system is 
\be
L  =  \frac{mR^2}{2} ( \dot\theta_1^2 + \dot\theta_2^2 + \dot\theta_3^2 )   - V(\theta,x,y) 
\ee
with potential energy 
\begin{align}
V(\theta,x,y) =   \frac{k}{2} \Bigl\{ &  (x - R\cos\theta_1)^2 + (y - R\sin\theta_1)^2 
	 + (x - R\cos\theta_2)^2 + (y - R\sin\theta_2)^2  \nono\\
	& +  (x - R\cos\theta_3)^2 + (y - R\sin\theta_3)^2  \Bigr\}  \ .
\end{align}
The conjugate momenta are 
\bse
\begin{align}
	p_i & \equiv \frac{\partial L}{\partial \dot\theta_i} = mR^2 \dot\theta_i \ ,\\
	p_x & \equiv \frac{\partial L}{\partial\dot x} = 0 \ ,\\
	p_y &  \equiv \frac{\partial L}{\partial\dot y} = 0 \ ,
\end{align}
\ese
where $i = 1$, $2$, and $3$. 
We have two primary constraints, 
\bse
\begin{align}
	\phi_1 & = p_x \ ,\\
	 \phi_2 & = p_y \  ,
\end{align}
\ese
and the primary Hamiltonian is 
\be
	H_P = \frac{1}{2mR^2} (p_1^2 + p_2^2 + p_3^2) + V(\theta,x,y) + \lambda_1 p_x + \lambda_2 p_y \ .
\ee
The consistency conditions $[\phi_a,H_P] = 0$  lead to the secondary constraints 
\bse
\begin{align}
	\psi_1 & = kR(\cos\theta_1 + \cos\theta_2 + \cos\theta_3) - 3kx \ ,\\
	\psi_2 & = kR(\sin\theta_1 + \sin\theta_2 + \sin\theta_3) - 3ky \ ,
\end{align}
\ese
and the conditions $[\psi_m,H_P] = 0$ yield restrictions on the Lagrange multipliers: 
\bse
\begin{align}
	\lambda_1 & = -\frac{1}{3mR} ( p_1\sin\theta_1 + p_2\sin\theta_2 + p_3\sin\theta_3) \ ,\\
	\lambda_2 & = \frac{1}{3mR} ( p_1\cos\theta_1 + p_2\cos\theta_2 + p_3\cos\theta_3) \ .
\end{align}
\ese
The secondary constraints tell us that the three springs connect at the point given by the average location of the three masses. This is required for the forces 
$\vec F_i = -k ( x - R\cos\theta_i, y - R\sin\theta_i)$ that the springs exert on the massless connection point to sum to zero. 

Inserting the results for the Lagrange multipliers into the primary Hamiltonian, we find the total Hamiltonian 
\begin{align}
	H_T =  & \frac{1}{2mR^2} (p_1^2 + p_2^2 + p_3^2) + V(\theta,x,y) 
	-\frac{p_x}{3mR} ( p_1\sin\theta_1 + p_2\sin\theta_2 + p_3\sin\theta_3)  \nono\\
	& + \frac{p_y}{3mR} ( p_1\cos\theta_1 + p_2\cos\theta_2 + p_3\cos\theta_3)  \ .
\end{align}
The constraints ${\cal C}^{(sc)}_\mu = \{\phi_1, \phi_2, \psi_1, \psi_2\}$ are second class, with Poisson bracket
\be
	{\cal M}_{\mu\nu} = [{\cal C}^{(sc)}_\mu, {\cal C}^{(sc)}_\nu ] = \begin{pmatrix} 0 & 0 & 3k & 0 \\ 0 & 0 & 0 & 3k \\
	-3k & 0 & 0 & 0 \\ 0 & -3k & 0 & 0 \end{pmatrix} \ .
\ee
We now construct the Dirac bracket as defined in Eq.~(\ref{DiracBracketsSC}). The brackets among the phase space 
variables include
\bse
\begin{align}
	[\theta_i,p_j]^* & =\delta_{ij} \ ,\\
	[x,p_i]^* &  = -\frac{R}{3} \sin\theta_i  \ ,\\
	[y,p_i]^* & = \frac{R}{3} \cos\theta_i \ ,
\end{align}
\ese
where $i$ and $j$ range over $1$, $2$, $3$. The remaining Dirac brackets vanish.  

We can use the constraints to eliminate four of the phase space variables; the natural choice is $x$, $y$, $p_x$ and $p_y$. From 
${\cal C}^{(sc)}_\mu = 0$ we find $p_x = p_y = 0$ and 
\bse\label{massesspringsringsxy}
\begin{align}
	x & = \frac{R}{3} (\cos\theta_1 + \cos\theta_2 + \cos\theta_3) \ ,\\
	y & = \frac{R}{3} (\sin\theta_1 + \sin\theta_2 + \sin\theta_3) \ .
\end{align}
\ese
The reduced Hamiltonian is
\be
	H_R = \frac{1}{2mR^2}(p_1^2 + p_2^2 + p_3^2) - \frac{kR^2}{3} \bigl( \cos(\theta_1 - \theta_2) + \cos(\theta_2 - \theta_3) + 
	\cos(\theta_3 - \theta_1) \bigr) + kR^2
\ee
From here we obtain the equations of motion for the angles, 
\be
	\dot\theta_i = [\theta_i,H_R]^* = p_i/(mR^2) \ ,
\ee
and their conjugate momenta, 
\bse
\begin{align}
	\dot p_1 & = [p_1,H_R]^* = \frac{kR^2}{3} \bigl( \sin(\theta_2 - \theta_1) + \sin(\theta_3 - \theta_1) \bigr) \ ,\\
	\dot p_2 & = [p_3,H_R]^* = \frac{kR^2}{3} \bigl( \sin(\theta_3 - \theta_2) + \sin(\theta_1 - \theta_2) \bigr) \ ,\\
	\dot p_3 & = [p_3,H_R]^* = \frac{kR^2}{3} \bigl( \sin(\theta_1 - \theta_3) + \sin(\theta_2 - \theta_3) \bigr) \ .
\end{align}
\ese
It is straightforward to solve these  equations numerically.  Equations (\ref{massesspringsringsxy})  then determine the coordinates 
$x$, $y$ as functions of time. 

%%%%%%%%%%%%%%%%%%%%%%%%%%%%%%%%%%%%%%%%
\subsection{Masses, Rods and Springs}\label{sec:massesrodssprings}
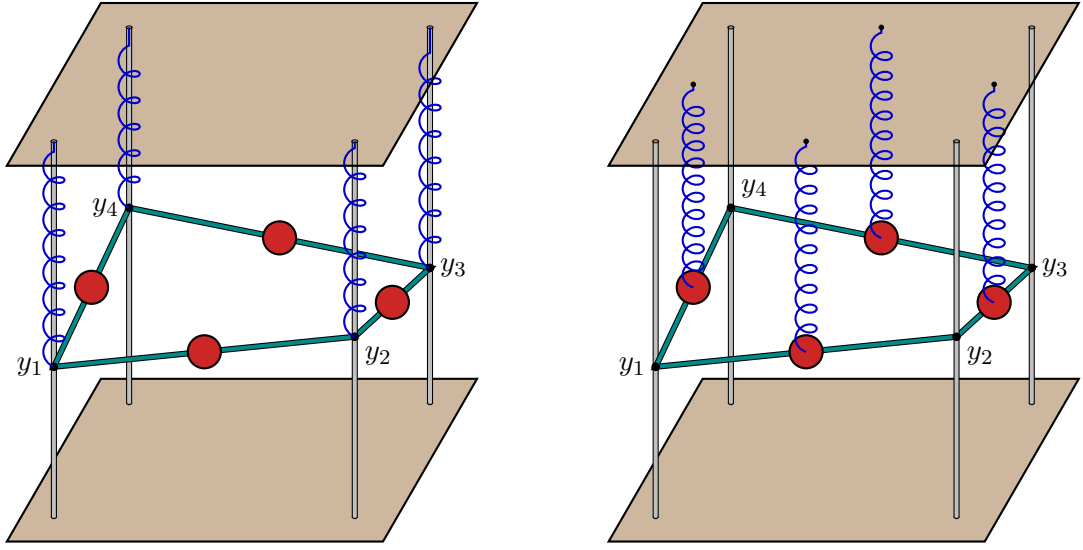
\begin{figure}[htb]
\centering
\begin{tikzpicture}
\begin{scope}[shift={(-4,0)}]
\def\xlen{5.0};
\def\ylen{2.5};
\def\angl{60};
\def\ch{5.0};
\tikzmath{
\dxp = 0.1*\xlen + 0.1*\ylen*cos(\angl);
\dxm = 0.1*\xlen - 0.1*\ylen*cos(\angl);
\dy = 0.15*\ylen*sin(\angl);
}
\coordinate (fl) at (\dxp,\dy);
\coordinate (fr) at (\xlen-\dxm,\dy);
\coordinate (bl) at ($(\angl:\ylen)+(\dxm,-\dy)$);
\coordinate (br) at ($(\xlen,0)+(\angl:\ylen)-(\dxp,\dy)$);
\coordinate (fldot) at ($(fl)+(0,2.0)$);
\coordinate (frdot) at ($(fr)+(0,2.4)$);
\coordinate (bldot) at ($(bl)+(0,2.6)$);
\coordinate (brdot) at ($(br)+(0,1.8)$);
\draw [draw=black, thick, fill=Bisque3]  (0,0)--(\xlen,0)-- +(\angl:\ylen)--(\angl:\ylen)--cycle;                               %square base
\draw [fill=Bisque3]  (0,\ch)--(\xlen,\ch) --+(\angl:\ylen)--($(\angl:\ylen) + (0,\ch)$) --cycle;    % square ceiling fill
\draw[double=gray!50!white, double distance=1.5pt] (fl) -- ($(fl)+(0,\ch)$);                                  %front left post (y1)
\draw []  ($(fl)+ (0,\ch)$)  ellipse (0.04 and 0.02) ;  
\centerarc[]($(fl) + (0,0.02)$)(200:340:0.04);
\draw[double=gray!50!white, double distance=1.5pt] (bl) -- ($(bl)+(0,\ch)$);                                           %back left post (y2)
\draw []  ($(bl)+ (0,\ch)$)  ellipse (0.04 and 0.02) ;  
\centerarc[]($(bl) + (0,0.02)$)(200:340:0.04);
\draw[double=gray!50!white, double distance=1.5pt] (br) -- ($(br)+(0,\ch)$);                                           %back right post (y3)
\draw []  ($(br)+ (0,\ch)$)  ellipse (0.04 and 0.02) ;  
\centerarc[]($(br) + (0,0.02)$)(200:340:0.04);
\draw [double = Cyan4, double distance=1.5pt]  (fldot)--(frdot)--(brdot)--(bldot)--cycle;
\draw[fill] (fldot) circle[radius=0.05];
\draw[fill] (bldot) circle[radius=0.05];
\draw[fill] (brdot) circle[radius=0.05];
\draw[draw=black, thick,  fill=Firebrick3] ($(fldot)!0.5!(frdot)$) circle[radius=0.22];
\draw[draw=black, thick,  fill=Firebrick3] ($(frdot)!0.5!(brdot)$) circle[radius=0.22];
\draw[draw=black, thick,  fill=Firebrick3] ($(brdot)!0.5!(bldot)$) circle[radius=0.22];
\draw[draw=black, thick,  fill=Firebrick3] ($(bldot)!0.5!(fldot)$) circle[radius=0.22];
\draw (fldot) node[left] {$y_1$};
\draw (frdot) node[below right] {$y_2$};
\draw (brdot) node[right] {$y_3$};
\draw (bldot) node[left] {$y_4$};
\draw[Blue3, snake=coil, segment amplitude=4pt, segment aspect=0.8, thick] (fldot) -- ($(fl) + (0,\ch)$);
\draw[Blue3, snake=coil, segment amplitude=4pt, segment aspect=0.8, thick] (bldot) -- ($(bl) + (0,\ch)$);
\draw[Blue3, snake=coil, segment amplitude=4pt, segment aspect=0.8, thick] (brdot) -- ($(br) + (0,\ch)$);
\draw[double=gray!50!white, double distance=1.5pt] (fr) -- ($(fr)+(0,\ch)$);                                           %front right post (y4)
\draw []  ($(fr)+ (0,\ch)$)  ellipse (0.04 and 0.02) ;  
\centerarc[]($(fr) + (0,0.02)$)(200:340:0.04);
\draw[fill] (frdot) circle[radius=0.05];
\draw[Blue3, snake=coil, segment amplitude=4pt, segment aspect=0.8, thick] (frdot) -- ($(fr) + (0,\ch)$);
\draw [draw=black, thick]  (0,\ch)--(\xlen,\ch) --+(\angl:\ylen)--($(\angl:\ylen) + (0,\ch)$) --cycle;    % square ceiling outline
\end{scope}
%%%%%%%%%%%%%%%%%
\begin{scope}[shift={(4,0)}]
\def\xlen{5.0};
\def\ylen{2.5};
\def\angl{60};
\def\ch{5.0};
\tikzmath{
\dxp = 0.1*\xlen + 0.1*\ylen*cos(\angl);
\dxm = 0.1*\xlen - 0.1*\ylen*cos(\angl);
\dy = 0.15*\ylen*sin(\angl);
}
\coordinate (fl) at (\dxp,\dy);
\coordinate (fr) at (\xlen-\dxm,\dy);
\coordinate (bl) at ($(\angl:\ylen)+(\dxm,-\dy)$);
\coordinate (br) at ($(\xlen,0)+(\angl:\ylen)-(\dxp,\dy)$);
\coordinate (fldot) at ($(fl)+(0,2.0)$);
\coordinate (frdot) at ($(fr)+(0,2.4)$);
\coordinate (bldot) at ($(bl)+(0,2.6)$);
\coordinate (brdot) at ($(br)+(0,1.8)$);
\draw [draw=black, thick, fill=Bisque3]  (0,0)--(\xlen,0)-- +(\angl:\ylen)--(\angl:\ylen)--cycle;                               %square base
\draw [fill = Bisque3]  (0,\ch)--(\xlen,\ch) --+(\angl:\ylen)--($(\angl:\ylen) + (0,\ch)$) --cycle;    % square ceiling fill
\draw[double=gray!50!white, double distance=1.5pt] (fl) -- ($(fl)+(0,\ch)$);                                  %front left post (y1)
\draw []  ($(fl)+ (0,\ch)$)  ellipse (0.04 and 0.02) ;  
\centerarc[]($(fl) + (0,0.02)$)(200:340:0.04);
\draw[double=gray!50!white, double distance=1.5pt] (bl) -- ($(bl)+(0,\ch)$);                                           %back left post (y2)
\draw []  ($(bl)+ (0,\ch)$)  ellipse (0.04 and 0.02) ;  
\centerarc[]($(bl) + (0,0.02)$)(200:340:0.04);
\draw[double=gray!50!white, double distance=1.5pt] (br) -- ($(br)+(0,\ch)$);                                           %back right post (y3)
\draw []  ($(br)+ (0,\ch)$)  ellipse (0.04 and 0.02) ;  
\centerarc[]($(br) + (0,0.02)$)(200:340:0.04);
\draw [double = Cyan4, double distance=1.5pt]  (fldot)--(frdot)--(brdot)--(bldot)--cycle;
\draw[fill] (fldot) circle[radius=0.05];
\draw[fill] (bldot) circle[radius=0.05];
\draw[fill] (brdot) circle[radius=0.05];

\draw[draw=black, thick,  fill=Firebrick3] ($(fldot)!0.5!(frdot)$) circle[radius=0.22];
\draw[draw=black, thick,  fill=Firebrick3] ($(frdot)!0.5!(brdot)$) circle[radius=0.22];
\draw[draw=black, thick,  fill=Firebrick3] ($(brdot)!0.5!(bldot)$) circle[radius=0.22];
\draw[draw=black, thick,  fill=Firebrick3] ($(bldot)!0.5!(fldot)$) circle[radius=0.22];
\draw (fldot) node[left] {$y_1$};
\draw (frdot) node[below right] {$y_2$};
\draw (brdot) node[right] {$y_3$};
\draw (bldot) node[above right] {$y_4$};
\draw[Blue3, snake=coil, segment amplitude=4pt, segment aspect=0.8, segment length=7.5, thick] ($(fldot)!0.5!(frdot)$) --  ($(fl)!0.5!(fr) + (0,\ch)$);
\draw[Blue3, snake=coil, segment amplitude=4pt, segment aspect=0.8, segment length=7, thick] ($(frdot)!0.5!(brdot)$) -- ($(fr)!0.5!(br) + (0,\ch)$);
\draw[Blue3, snake=coil, segment amplitude=4pt, segment aspect=0.8, segment length=7.5, thick] ($(brdot)!0.5!(bldot)$) -- ($(bl)!0.5!(br) + (0,\ch)$);
\draw[Blue3, snake=coil, segment amplitude=4pt, segment aspect=0.8, segment length=6.5, thick] ($(bldot)!0.5!(fldot)$) -- ($(bl)!0.5!(fl) + (0,\ch)$);
\draw[fill] ($(fl)!0.5!(fr) + (0,\ch)$) circle[radius=0.03];
\draw[fill] ($(fr)!0.5!(br) + (0,\ch)$) circle[radius=0.03];
\draw[fill] ($(bl)!0.5!(br) + (0,\ch)$) circle[radius=0.03];
\draw[fill] ($(bl)!0.5!(fl) + (0,\ch)$) circle[radius=0.03];
\draw[double=gray!50!white, double distance=1.5pt] (fr) -- ($(fr)+(0,\ch)$);                                           %front right post (y4)
\draw []  ($(fr)+ (0,\ch)$)  ellipse (0.04 and 0.02) ;  
\centerarc[]($(fr) + (0,0.02)$)(200:340:0.04);
\draw[fill] (frdot) circle[radius=0.05];
\draw [draw=black, thick]  (0,\ch)--(\xlen,\ch) --+(\angl:\ylen)--($(\angl:\ylen) + (0,\ch)$) --cycle;    % square ceiling outline
\end{scope}
\end{tikzpicture}
\caption{Four massless, freely extensible  rods are connected to each other at the corner posts. The rods slide freely along the posts.  
In the left figure  springs are attached to the corners. In the right figure springs are attached to the masses.}
\label{fig:massesrodssprings}
\end{figure}

The system shown in Fig.~\ref{fig:massesrodssprings}  consists of 
four masses fixed at the midpoints of four massless, freely extensible rods. A freely extensible rod is rigid in transverse directions but does not have a fixed length.\cite{extensiblerods} 
That is, the rods can expand or contract as needed to span the distance between the vertical posts.
Figure \ref{fig:massesrodssprings} shows  two versions of this system. In the left figure,  springs are attached to the connection points between the rods 
(the ``corners" with coordinates $y_1$ through $y_4$). In the right figure  the springs are attached to the masses. 

The Lagrangian for this system is
\be
	L = \frac{m}{2} \left[ \left(\frac{\dot y_1 + \dot y_2}{2} \right)^2 + \left(\frac{\dot y_2 + \dot y_3}{2} \right)^2 
	+ \left(\frac{\dot y_3 + \dot y_4}{2}\right)^2 + \left(\frac{\dot y_4 + \dot y_1}{2} \right)^2 \right] - V(y) \ ,
\ee
with potential energy
\be\label{Vendsprings}
	V(y)   =  mg \left[  y_1 + y_2 + y_3 + y_4 \right] 
	 + \frac{k}{2} \left[ (a - y_1)^2 + (a - y_2)^2 + (a - y_3)^2 + (a - y_4)^2 \right]
\ee
when the springs are attached to the corners, and 
\begin{align}\label{Vmidsprings}
	V(y) & =  mg \left[  y_1 + y_2 + y_3 + y_4 \right]   \nono\\
	& + \frac{k}{8} \left[ (2a - y_1 - y_2)^2 + (2a - y_2 - y_3)^2 + (2a - y_3 - y_4)^2 + (2a - y_4 - y_1)^2 \right] 
\end{align}
when the springs are attached to the masses. 
Here, $k$ denotes the spring constant and $a=h-\ell$, where $h$ is the height of the ceiling and $\ell$ is the 
relaxed length of each spring. 

The momenta are 
\bse
\begin{align}
	p_1 & = \frac{m}{4} ( \dot y_4 + 2\dot y_1 + \dot y_2) \ , \\
	p_2 & = \frac{m}{4} ( \dot y_1 + 2\dot y_2 + \dot y_3) \ ,\\
	p_3 & = \frac{m}{4} ( \dot y_2 + 2\dot y_3 + \dot y_4) \ ,\\
	p_4 & = \frac{m}{4} ( \dot y_3 + 2\dot y_4 + \dot y_1) \ ,
\end{align}
\ese
which yield the single primary constraint 
\be
	\phi = p_1 - p_2 + p_3 - p_4 \ .
\ee
The  canonical Hamiltonian is given by 
\be
	H_C = \frac{1}{2m} \left[ \frac{5}{2} (p_1^2 + p_2^2 + p_3^2 + p_4^2 )  - 2(p_1p_2 + p_2p_3 + p_3p_4 + p_4p_1) + p_1p_3 + p_2 p_4 \right] + V(y) \ ,
\ee
and the primary Hamiltonian is $H_P = H_C + \lambda \phi$. 

Focus on the case shown on the left of  Fig.~\ref{fig:massesrodssprings}, with springs attached to the corners and the potential energy  
of Eq.~(\ref{Vendsprings}). 
The consistency conditions yield a secondary constraint 
\be
	\psi = -k(y_1 - y_2 + y_3 - y_4) \ ,
\ee
and the restriction 
\be
	\lambda = -\frac{5}{4m} ( p_1 - p_2 + p_3 - p_4)
\ee
on the Lagrange multiplier.  The Lagrange multiplier can be simplified to $\lambda = 0$ using the constraint $\phi = 0$.   

The secondary constraint is interesting because it places a restriction on the configuration of the system that might not be obvious. 
One can imagine  moving any one of the corners, by hand,  independently of the others. The mechanical arrangement of rods and posts do not place any restrictions 
on the $y$ values. But as a dynamical 
system, the $y$'s must obey $y_1 - y_2 + y_3 - y_4 = 0$. This is because the $y$'s can be changed while keeping the masses in place. For example, if $y_1$ and $y_3$ are increased 
by some amount $\delta y$, while $y_2$ and $y_4$ are decreased by the same amount $\delta y$, the masses remain unmoved. 
There is no inertial resistance to this type of motion. As a result, the springs 
can instantly ``snap" the corners into the preferred configuration satisfying  $y_1 - y_2 + y_3 - y_4 = 0$. This is the configuration 
that minimizes the potential (\ref{Vendsprings}) while keeping the locations of the masses fixed. 

In this example the total Hamiltonian $H_T$ coincides with the 
canonical Hamiltonian $H_C$. The constraints $\phi$, $\psi$ are second class, so we can use them to eliminate two of the phase space variables. For example, let 
\bse\label{y4p4eqns}
\begin{align}
	y_4 & =  y_1 - y_2 + y_3 \ ,\\
	p_4 & = p_1 - p_2 + p_3 \ .
\end{align}
\ese
Then the reduced Hamiltonian becomes  
\begin{align}
	H_R & = \frac{1}{2m} \left[ 3 (p_1^2 + p_2^2 + p_3^2) -4(p_1 p_2 + p_2p_3) + 2p_1 p_3 \right] + 2gm(y_1 + y_3)  \nono\\
	& + k\left[ y_1^2 + y_2^2 + y_3^2 - y_1 y_2 + y_1 y_3 - y_2 y_3 - 2a(y_1 + y_3 - a) \right] \ .
\end{align}
The equations of motion are obtained from $H_R$ and the Dirac bracket. Writing these as a system of second order equations for the 
coordinates, we find 
\bse
\begin{align}
	\ddot y_1 & = -g - \frac{k}{2m} ( 3 y_1 - y_3 - 2a) \ ,\\
	\ddot y_2 & = -g - \frac{k}{2m} ( 4 y_2 - y_1 - y_3 - 2a) \ ,\\
	\ddot y_3 & = -g - \frac{k}{2m} ( 3 y_3 - y_1 - 2a) \ .
\end{align}
\ese
This simple system of linear equations can be solved analytically for $y_1$, $y_2$ and $y_3$ as functions of time $t$. 
The remaining variable $y_4(t)$ is determined from Eq.~(\ref{y4p4eqns}a).  

The general solution is a linear combination 
of three harmonic modes: (i) the rods remain horizontal ($y_1 = y_2 = y_3 = y_4$)  and oscillate up and down  
with frequency $\sqrt{k/m}$; (ii) the rod between $y_1$ and $y_2$ and  
the rod between $y_3$ and $y_4$ remain horizontal as they oscillate $180^\circ$ out of phase with frequency  $\sqrt{2k/m}$; (iii) the rod between  $y_2$ and $y_3$ and  
the rod between $y_4$ and $y_1$ remain horizontal as they oscillate $180^\circ$ out of phase with frequency  $\sqrt{2k/m}$. 

Now specialize to the system shown on the right side of Fig.~\ref{fig:massesrodssprings}, with springs attached to the masses and the potential energy  
of Eq.~(\ref{Vmidsprings}).  
In this case the time derivative of $\phi$ vanishes, $[\phi,H_P] = 0$, so there are no secondary constraints and the Lagrange multiplier $\lambda$ 
is unrestricted. The total Hamiltonian $H_T$ coincides with the primary Hamiltonian $H_P$. 

Since $\phi$ is the only constraint, it is necessarily first class and it generates a gauge transformation. Explicitly, let $G = \epsilon\phi$ where 
$\epsilon$ is an arbitrary function of time. The gauge transformation of any phase space function $F(q,p)$ is determined by the Poisson bracket  of $F$ with the 
gauge generator $G$; that is, $\delta F = [F,G]$. 
For the phase space coordinates, we have
$\delta y_1  = \epsilon$,  $\delta y_2  = -\epsilon$, $\delta y_3   = \epsilon$,  $\delta y_4  = -\epsilon$ and $\delta p_i = 0$ (with $i = 1,\ldots,4$). 
This describes a change in the $q$'s and $p$'s for which the masses don't move and the spring lengths  don't change. In other words, the physical state of the 
system is unchanged. 

Observables are phase space functions that are gauge invariant. Observables include the locations of the masses, namely, 
$x_{12} \equiv (y_1 + y_2)/2$,  $x_{23} \equiv (y_2 + y_3)/2$,  $x_{34}\equiv (y_3 + y_4)/2$ and  $x_{41}\equiv (y_4 + y_1)/2$, and the momenta $p_i$. We can identify the physical meaning of the $p$'s by computing 
the time derivatives of the masses' locations using the total Hamiltonian. This shows that $p_i = mv_i$, where $v_i$  is the average of the  velocities of the 
two masses adjacent to the corner $y_i$.  

From this simple example we see that each physical state of the system is described by a curve in phase space, a curve defined by the transformation  $\delta F =  \epsilon [F,\phi]$. 
These  curves are called ``gauge orbits". 

We can select a single point on each gauge orbit to represent the physical state of the system. We do this by choosing a gauge condition $\chi(q,p) = 0$ 
such that $\chi$ and $\phi$, together, form a set of  second class constraints.  For example, we can require the rod between $y_1$ and $y_2$ to remain horizontal by choosing 
\be\label{gauge2}
	\chi = y_2 - y_1 \ .
\ee
Since  $[\chi,\phi]  \ne 0$, the set ${\cal C}^{(all)}_M = \{ \chi , \phi \} $ is indeed second class.  
The Dirac bracket is constructed as in Eq.~(\ref{DiracBracketsALL}), and we can  use the constraints to eliminate two of the variables, say, $y_1$ and $p_1$. 
The fully reduced Hamiltonian is 
\be
	H_{FR} = \frac{1}{2m} \left[  3(p_2^2 + p_4^2 ) + 4(p_3^2 - p_2p_3 - p_4 p_3) + 2p_2 p_4 \right] 
	+ V(y)\bigr|_{y_1 = y_2}
\ee
where $V(y)\bigr|_{y_1 = y_2}$ is
 the potential energy of Eq.~(\ref{Vmidsprings}) evaluated at $y_1 = y_2$. The nonzero Dirac brackets among the 
remaining variables are 
\bse
\begin{align}
	[y_3, p_3]^*  = [y_4, p_4]^* & = 1 \ ,\\
	[y_2,p_2]^*  = [y_3,p_2]^* = -[y_4,p_2]^* & = 1/2 \ ,
\end{align}
\ese
and the equations of motion $\dot F = [F,H_{FR}]^*$ are  
\bse
\begin{align}
	\dot y_2 & = \frac{1}{2m} ( 3p_2 - 2p_3 + p_4 ) \ ,\\
	\dot y_3 & = \frac{1}{2m} (-p_2 + 6 p_3 - 3p_4) \ ,\\
	\dot y_4 & = \frac{1}{2m}( -p_2 - 2p_3 + 5p_4) \ ,\\
	\dot p_2 & = ak-mg - \frac{k}{4} (3 y_2 + y_3) \ ,\\
	\dot p_3 & = ak - mg - \frac{k}{4}(y_2 +2y_3 + y_4)  \ ,\\
	\dot p_4 & = ak-mg - \frac{k}{4} ( y_2 + y_3 + 2y_4) \ .
\end{align}
\ese
These results imply  $\ddot y_i = ak/m - g - (k/m)y_i $ where $i = 2,3,4$. Thus, each of the three corners $y_2$, $y_3$, and $y_4$ independently execute simple 
harmonic motion with frequency $\sqrt{k/m}$ about their equilibrium positions $a - mg/k$. The corner $y_1$ moves in sync with $y_2$, due to the gauge condition $\chi  = y_2 - y_1 = 0$. 

There are  other interesting gauge choices. For example, we can freeze the corner $y_4$ by letting  
\be\label{gauge1}
	\chi = y_4 - a + mg/k \ .
\ee
After constructing the Dirac bracket,  eliminating the variables $x_4$ and $p_4$ and reducing the Hamiltonian,  we find 
the equations $\ddot y_i = ak/m - g - (k/m)y_i $ for $i = 1,2,3$.   The corners $y_1$, $y_2$, and $y_3$ independently execute simple 
harmonic motion with frequency $\sqrt{k/m}$, while $y_4$ remains fixed. 

The evolution of the observables is, of course, independent of the gauge choice.  In particular, the motions of the masses are the same whether we choose  (\ref{gauge2}) or (\ref{gauge1}). 
It is not difficult to see that each of the four masses executes simple harmonic motion with frequency $\sqrt{k/m}$. However, the amplitudes and 
phases are not independent---they must satisfy 
$x_{12} + x_{34} = x_{23} + x_{41}$.   This relationship follows from the definitions $x_{12} \equiv (y_1 + y_2)/2$, {\em etc}.

%%%%%%%%%%%%%%%%%%%%%%%%%%%%%%%%%%%
\subsection{Pairs of  pulleys}\label{sec:pairsofpulleys}
Figure \ref{fig:pairsofpulleys} shows three pairs of massless pulleys. Each pair consists of a fixed lower pulley and an upper pulley attached to a mass and a spring. 
A cord runs over and under the pulleys, as shown, with the left end attached to the right end. That is, the left and right sides of the figure are ``periodically identified" 
so that the cord forms a single continuous loop. (This can be achieved in three dimensions by 
attaching the springs and the bottom pulleys to circular supports.) 
Note that we can construct such a system using any number of pairs of pulleys. We will focus on the version with three pairs. 
\def\rad{0.6}
	\def\rshift{4*\rad}
	\def\lshift{-4*\rad}
	\def\up{1.0}
	\def\upr{1.5}
	\def\upl{2.0}
	\def\cht{4.5} % ceiling height
	 \def\flr{-2.3} % floor height
	\def\lw{1.5} 
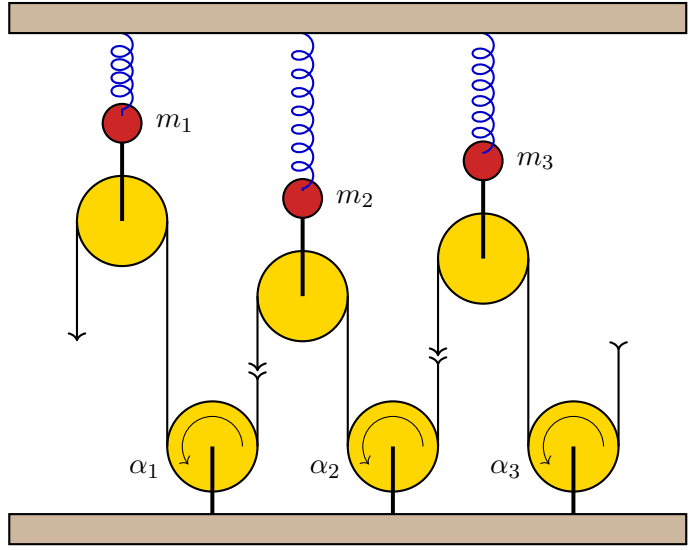
\begin{figure}
\centering
\begin{tikzpicture}
		% Middle pair
	\node (a) at (2*\rad,-1) {};
	\node (b) at (\rad,-1) {};
	\node (c) at (0.2,-1.3) {}; 
	\draw [thick, draw=black, fill=Gold1]  (\rad,-1)  circle (\rad);
	\node at (\rad-0.9, -1.3)  [] {$\alpha_2$};
	\draw  pic[draw, ->, angle eccentricity=1.2, angle radius=0.4cm]  {angle=a--b--c};
	\draw [line width=\lw] (\rad,-1) -- (\rad,\flr + 0.4);
	\draw [thick, draw=black, fill=Gold1]  (-\rad,1) circle (\rad);
	\draw[line width=\lw] (-\rad,1) -- (-\rad, 2.3);
	\draw [draw=black, thick,  fill=Firebrick3] (-\rad,2.3) circle (0.43*\rad) ;   
	\node at (-\rad + 0.7, 2.3) [] {$m_2$};  
	\draw[Blue3, snake=coil, segment amplitude=4pt, segment aspect=0.8, segment length=7,  thick]  (-\rad, \cht) -- (-\rad,2.4) ;
	\draw [thick] (0,-1) -- (0,\up);
	\draw [thick, ->] (-2*\rad,\up) -- (-2*\rad,0);
	\draw [thick, -<] (2*\rad,-1) -- (2*\rad,0.2);
	% Right pair
	\node (ar) at (\rshift +2*\rad,-1) {};
	\node (br) at (\rshift + \rad,-1) {};
	\node (cr) at (\rshift + 0.2,-1.3) {}; 
	\draw [thick, draw=black, fill=Gold1]  (\rshift+\rad,-1) circle (\rad);
	\node at (\rshift+\rad-0.9, -1.3) [] {$\alpha_3$};
	\draw  pic[draw, ->, angle eccentricity=1.2, angle radius=0.4cm]  {angle=ar--br--cr};
	\draw [line width=\lw] (\rshift+\rad,-1) -- (\rshift+\rad,\flr + 0.4);
	\draw [thick, draw=black, fill=Gold1]  (\rshift-\rad, \upr) circle (\rad);
	\draw[line width=\lw] (\rshift-\rad,\upr) -- (\rshift-\rad, \upr + 1.3);
	\draw[draw=black, thick,  fill=Firebrick3] (\rshift-\rad,\upr+1.3) circle (0.43*\rad) ;
	\node at (\rshift-\rad + 0.7, \upr+1.3) [] {$m_3$};  
	\draw[Blue3, snake=coil, segment amplitude=4pt, segment aspect=0.8, segment length=6,  thick]  (\rshift-\rad, \cht) -- (\rshift-\rad,\upr+1.4) ;
	\draw [thick] (\rshift,-1) -- (\rshift,\upr);
	\draw [thick, ->] (\rshift-2*\rad,\upr) -- (\rshift-2*\rad,0.2);
	\draw [thick, -<] (\rshift+2*\rad,-1) -- (\rshift+2*\rad,0.4);
	% Left pair
	\node (al) at (\lshift +2*\rad,-1) {};
	\node (bl) at (\lshift + \rad,-1) {};
	\node (cl) at (\lshift + 0.2,-1.3) {}; 
	\draw [thick, draw=black, fill=Gold1]  (\lshift+\rad,-1) circle (\rad);
	\node at (\lshift+\rad-0.9, -1.3) [] {$\alpha_1$};
	\draw  pic[draw, ->, angle eccentricity=1.2, angle radius=0.4cm]  {angle=al--bl--cl};
	\draw [line width=\lw] (\lshift+\rad,-1) -- (\lshift+\rad,\flr + 0.4);
	\draw [thick, draw=black, fill=Gold1]  (\lshift-\rad,\upl) circle (\rad);
	\draw[line width=\lw] (\lshift-\rad,\upl) -- (\lshift-\rad, \upl + 1.3);
	\draw [draw=black, thick,  fill=Firebrick3] (\lshift-\rad,\upl+1.3) circle (0.43*\rad);
	\node at (\lshift-\rad + 0.7, \upl+1.3) [] {$m_1$};  
	\draw[Blue3, snake=coil, segment amplitude=4pt, segment aspect=0.8, segment length=5,  thick]  (\lshift-\rad, \cht) -- (\lshift-\rad,\upl+1.4) ;
	\draw [thick] (\lshift,-1) -- (\lshift,\upl);
	\draw [thick, ->] (\lshift-2*\rad,\upl) -- (\lshift-2*\rad,0.4);
	\draw [thick, -<] (\lshift+2*\rad,-1) -- (\lshift+2*\rad,0);
	%  Ceiling and Floor
	\draw [draw=black, thick, fill=Bisque3]  (-4.5,\cht)--(4.5,\cht)--(4.5,\cht+0.4)--(-4.5,\cht+0.4)--cycle;
	\draw [draw=black, thick, fill=Bisque3]  (-4.5,\flr)--(4.5,\flr)--(4.5,\flr+0.4)--(-4.5,\flr+0.4)--cycle;
   	\end{tikzpicture}
\caption{A loop of cord winds through three pairs of pulleys; the left and right ends of the figure are identified.  Masses and springs are attached to each of the upper pulleys. This system can be extended to any number of pairs of pulleys.}
\label{fig:pairsofpulleys}
\end{figure}

The coordinates for this system 
are the angles $\alpha_1$, $\alpha_2$, $\alpha_3$ of the lower, fixed pulleys. Consider the height of mass $m_2$. If the angle $\alpha_1$ increases by $\delta\alpha_1$, the height of $m_2$ 
increases by $R\, \delta\alpha_1/2$ where $R$ is the radius of the lower pulley. If the angle $\alpha_2$ increases by 
$\delta\alpha_2$, the height of $m_2$ decreases by $R\, \delta\alpha_2/2$. Thus we see that the height of $m_2$ is $h_2 = R(\alpha_1 - \alpha_2)/2  + c$, where $c$ is a constant. Likewise, the height of $m_1$ is $h_1 = R(\alpha_3 - \alpha_1)/2 + c$ and the height of $m_3$ is $h_3 = R(\alpha_2 - \alpha_3)/2 + c$.%\cite{constantc}  

Let $m_1 = m_2 = m_3$ and use the common notation $m$ for each mass. The kinetic energy for this system is $T = (m/2)(\dot h_1^2 + \dot h_2^2 + \dot h_3^2)$, or 
\be
	T = \frac{mR^2}{8} \left[ (\dot\alpha_1  - \dot\alpha_2)^2 + (\dot \alpha_2 - \dot\alpha_3)^2 + (\dot\alpha_3 - \dot\alpha_1)^2 \right] \ .
\ee
The gravitational potential energy  $mg(h_1 + h_2 + h_3)$ is simply a constant. The spring potential energy is 
$ (k/2) [(a - h_1)^2 + (a - h_2)^2 + (a - h_3)^2 ]$,  where the  constant $a$ depends on the height of the ceiling and the relaxed length of each spring. To within an additive 
constant, the total potential energy is
\be\label{Vofalpha}
	V(\alpha) = \frac{k R^2}{8} \left[ (\alpha_1 - \alpha_2)^2 + (\alpha_2 - \alpha_3)^2 + (\alpha_3 - \alpha_1)^2 \right] \ .
\ee
As usual the Lagrangian is $L = T-V$.

The momenta for this system are 
\bse
\begin{align}
	p_1 & = \frac{mR^2}{4} (2\dot\alpha_1 - \dot\alpha_2 - \dot\alpha_3)  \ ,\\
	p_2 & = \frac{mR^2}{4} (2\dot\alpha_2 - \dot\alpha_3 - \dot\alpha_1) \ , \\
	p_3 & = \frac{mR^2}{4} (2\dot\alpha_3 - \dot\alpha_1 - \dot\alpha_2)  \ .
\end{align}
\ese
The matrix of second derivatives $\partial^2L/\partial\dot\alpha_i \partial\dot\alpha_j$ has rank $2$ and there is  one primary constraint: 
\be
	\phi = p_1 + p_2 + p_3 \ .
\ee
The canonical Hamiltonian can be written as 
\be
	H_C = \frac{4}{3mR^2} ( p_1^2 + p_1 p_2  + p_2^2) + V(\alpha)
\ee
where $V(\alpha)$ is given in Eq.~(\ref{Vofalpha}). The primary Hamiltonian is $H_P = H_C + \lambda\phi$. 

The Poisson bracket  $[\phi,H_P] $ vanishes identically, so the consistency condition does not lead to any further constraints and does not restrict the Lagrange 
multiplier.  The primary constraint $\phi$ is first class.  We see that this is a gauge theory with gauge generator $G = \epsilon\phi$.  

Under a gauge transformation the phase space coordinates transform as 
\bse
\begin{align}
	\delta \alpha_i & = [\alpha_i,G] = \epsilon \ , \\
	\delta p_i & = [p_i,G] = 0 \ , 
\end{align}
\ese
for $i = 1,2,3$.  
Physically, the gauge freedom arises because the pulleys can rotate by equal amounts  ($\delta\alpha_1 = \delta\alpha_2 = \delta\alpha_3$),  causing the cord to cycle 
through the system while leaving each mass  fixed in place.  The observables for this system include the differences, $\alpha_3 - \alpha_1$, $\alpha_1 - \alpha_2$, $\alpha_2 - \alpha_3$, which are 
proportional to the heights $h_1$, $h_2$, $h_3$ of the masses. The observables also include the momenta $p_i$. In physical terms, these are given by 
\bse
\begin{align}
	p_1 & = mR(v_1 - v_3)/2 \ ,\\
	p_2 & = mR(v_2 - v_1)/2 \ ,\\
	p_3 & = mR(v_3 - v_2)/2 \ ,
\end{align}
\ese
where $v_1$, $v_2$, and $v_3$ are the velocities of the three masses. 

Let us fix the gauge with the condition $\chi = 0$ where
\be
	\chi = \alpha_1 + \alpha_2 + \alpha_3 \ .
\ee
The set $\{\chi$, $\phi\}$ is second class, and the nonzero Dirac brackets are $[\alpha_i,p_j]^* = 2/3$ for $i = j$ 
and $[\alpha_i,p_j]^* = -1/3$ for $i\ne j$. We can use $\chi=0$ and $\phi=0$ to eliminate two of the variables, say, $\alpha_3$ and $p_1$. 
Then the fully reduced Hamiltonian is 
\be
	H_{FR} = \frac{4}{3mR^2}\left( p_2^2 + p_2 p_3 + p_3^2 \right)  + \frac{3kR^2}{4} \left( \alpha_1^2 + \alpha_1\alpha_2 + \alpha_2^2 \right) \ .
\ee
The equations of motion for the remaining variables are 
\bse
\begin{align}
	\dot \alpha_1 & = -\frac{4}{3mR^2} (p_2 + p_3) \ ,\\
	\dot\alpha_2 & = \frac{4}{3mR^2} \, p_2 \ ,\\
	\dot p_2 & = -\frac{3kR^2}{4} \,  \alpha_2  \ ,\\
	\dot p_3 & = \frac{3kR^2}{4} (\alpha_1 + \alpha_2) \ .
\end{align}
\ese
The resulting second order equations for the angles are $\ddot\alpha_1 = -(k/m)\alpha_1$ and $\ddot\alpha_2 = -(k/m)\alpha_2$. Thus, 
the pulleys $\alpha_1$ and $\alpha_2$ execute independent simple harmonic motion with angular frequency $\sqrt{k/m}$. The third angle is 
determined from the gauge condition as $\alpha_3 = -\alpha_1 - \alpha_2$. 

The observables for this system include the heights $h_i$ of the masses. Each mass executes simple harmonic motion with frequency $\sqrt{k/m}$, subject to the 
restriction $h_1 + h_2 + h_3 = {\rm const}$. The restriction follows from the relations $h_1 = R(\alpha_3 - \alpha_1)/2 + c$, {\em etc}.

%%%%%%%%%%%%%%%%%%%%%%%%%%%%%%%%%%%%%
\section{Exercises for the Reader}\label{section:more}
The following problems are left as exercises for the reader.  Solutions can be found in the supplementary material.\cite{supplementwebsite} 

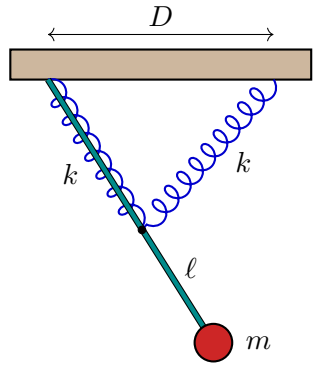
\begin{figure}[htb] 
\centering
\begin{tikzpicture}[]
      \draw [draw=black, thick,  fill=Bisque3] (-2,0)--(2,0)--(2,0.4)--(-2,0.4)--cycle;
      \draw [<->] (-1.5,0.6)--(1.5,0.6);
      \node at (0,0.85) {$D$};
    % springs
    \draw[Blue3, snake=coil,segment amplitude=4pt, segment aspect=0.8, segment length=8, thick] (-1.5,0) -- (-0.25,-2);
    \draw[Blue3, snake=coil,segment amplitude=4pt, segment aspect=0.8, segment length=7, thick] (1.5,0) -- (-0.25,-2);
    \node at (-1.2,-1.25) {$k$};
    \node at (1.1,-1.1) {$k$}; 
    % pendulum
    \draw [double=Cyan4, double distance=1.7pt] (-1.5,0) -- (0.7,-3.5);
    \draw [draw=black, thick,  fill=Firebrick3] (0.7,-3.5) circle (0.25);
    \node at (1.3, -3.5) {$m$};
    \node at (0.4,-2.5) {$\ell$};
    \draw [fill] (-0.25,-2) circle (0.05);
\end{tikzpicture}
\caption{A pendulum with two springs. One spring is wound around the pendulum rod. }
\label{fig:exercise1}
\end{figure}

1. A pendulum of mass $m$ and length $\ell$ hangs from the ceiling, as shown 
in Fig.~\ref{fig:exercise1}.  Two massless springs are attached to the ceiling, a distance $D$ apart, with spring $\#1$ wound around the pendulum rod. The two springs are 
attached to each other.   Let each spring 
have stiffness $k$ and a relaxed length of zero.   Use the angle of the pendulum and the length of spring $\#1$ as generalized coordinates. 

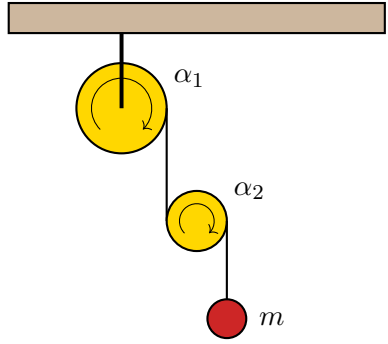
\begin{figure}[htb]
\centering
\begin{tikzpicture}
	\draw [draw=black, thick, fill=Bisque3]  (-2.5,0)--(2.5,0)--(2.5,0.4)--(-2.5,0.4)--cycle;
	\draw [thick, draw=black, fill=Gold1]  (-1,-1)  circle (0.6);
	\node at (-0.1, -0.6)  [] {$\alpha_1$};
	\node (d) at (0,-2) {};
	\node (e) at (-1,-1) {};
	\node (f) at (-2,-2) {};
	\draw  pic[draw, <-, angle eccentricity=1.2, angle radius=0.4cm]  {angle=d--e--f};
	\draw [line width=\lw] (-1,-1) -- (-1,0);
	\draw [thick] (-0.4,-1) -- (-0.4,-2.5);
	\draw [thick, draw=black, fill=Gold1]  (0,-2.5)  circle (0.4);
	\node at (0.7, -2.1)  [] {$\alpha_2$};
	\node (dd) at (1,-3.5) {};
	\node (ee) at (0,-2.5) {};
	\node (ff) at (-1,-3.5) {};
	\draw  pic[draw, <-, angle eccentricity=1.2, angle radius=0.23cm]  {angle=dd--ee--ff};
	\draw [thick] (0.4,-2.5) -- (0.4,-3.8);
	\draw [draw=black, thick,  fill=Firebrick3] (0.4,-3.8) circle (0.43*\rad) ;   
	\node at (1.0,-3.8) [] {$m$};  
\end{tikzpicture}
\caption{A mass $m$ hanging from a series of pulleys. The radii of the two pulleys can be different.}
\label{fig:exercise2}
\end{figure}

2. Two massless, frictionless pulleys are arranged as shown in Fig.~\ref{fig:exercise2}. The axis of the upper pulley is fixed, while the lower pulley is free to move vertically. The mass $m$ is also restricted  to move vertically. Note the direction in which the 
cords are wound around the pulleys. Use the orientation angles $\alpha_1$ and $\alpha_2$  as generalized coordinates. 

%%%%%%%%%%%%%%%%%%%%%%%%
\section{Conclusions}\label{sec:conclude}
Classical mechanics textbooks typically avoid singular Lagrangians by fiat. For many popular books, such as {\em Classical Mechanics} by Goldstein\cite{Goldstein} and 
{\em Mechanics} by Landau 
and Lifschitz,\cite{LandauLifshitz} this is understandable since these were written before the work of Dirac and Bergmann. 
But today we need not limit our attention to nonsingular systems. 
The Dirac--Bergmann algorithm is a natural extension of the standard Lagrangian and Hamiltonian formalism, and is not overly difficult to apply. It allows us to 
analyze interesting singular systems  and creates a closer link to modern field theories with gauge freedom. 

%%%%%%%%%%%
\begin{acknowledgments}
I thank the referees for insights and suggestions.  The author has no conflicts to disclose. 
\end{acknowledgments}

%%%%%%%%%%%%%

%%%%%%%%%%%%%%%

\begin{thebibliography}{99}
\bibitem{Lquadratic} 
If the Lagrangian is quadratic in the velocities, as is typical for systems in classical mechanics, it has the form 
$L(q,\dot q) = \dot q_i L_{ij} \dot q_j/2 + L_i \dot q_i + L_0$ where
$L_{ij}$, $L_i$, and $L_0$ depend only on the coordinates $q_i$. The conjugate momenta are then  $p_i = L_{ij} \dot q_j + L_i$. Let $V^i$ denote a vector in the 
null space of $L_{ij}$, so that  $V^i L_{ij} = 0$. If follow that $\phi = V^i p_i - V^i L_i = 0$ is a constraint on the $q$'s and $p$'s.
\bibitem{Diracconjecturecomment} If one accepts the Dirac conjecture, then all linear combinations of first class constraints  generate gauge transformations. 
If the Lagrangian is only invariant under particular linear combinations of first class constraints, then the Lagrangian theory can be viewed as a partially gauge--fixed theory. In other 
words, the gauge symmetry expressed in the Lagrangian theory is a subset of the gauge freedom of the extended Hamiltonian theory.
\bibitem{pendulumtwospringscomment} Here we assume $y<0$ and $-\pi/2 \le \theta \le \pi$, which implies $Y = y - \ell\cos\theta < 0$.
\bibitem{extensiblerods} A freely extensible rod can be approximated by  
a series of fixed--length rods and cylinders, where the rods slide through the cylinders without friction. A vertical guide can be used to keep the mass at the 
midpoint between the vertical posts.
\bibitem{supplementwebsite} URL to be inserted by AIPP.
\end{thebibliography}
\end{document}